\documentclass[twocolumn]{aastex631}

\newcommand{\teff}{T_\mathrm{eff}}
\newcommand{\logg}{\log g}
\newcommand{\abun}{[\mathrm{M}/\mathrm{H}]}

\received{}
\revised{}
\accepted{}

\submitjournal{ApJS}

\begin{document}

\title{Physics of Eclipsing Binaries. VI. Hot, compact stars}
\shorttitle{PHOEBE VI: Hot, compact stars}
\shortauthors{Jones et al.}

\correspondingauthor{David Jones}
\email{djones@iac.es}

\author[0000-0003-3947-5946]{David Jones}
\affiliation{Instituto de Astrof\'isica de Canarias,
E-38205 La Laguna,
Tenerife, 
Spain}
\affiliation{Departamento de Astrof\'isica,
Universidad de La Laguna,
E-38206 La Laguna,
Tenerife, 
Spain}

\author[0000-0002-0119-7883]{Nicole Reindl}
\affiliation{Landessternwarte Heidelberg, Zentrum f\"ur Astronomie, Ruprecht-Karls-Universit\"at, K\"onigstuhl 12, 69117, Heidelberg, Germany}

\author[0000-0002-5442-8550]{Kyle E. Conroy}
\affiliation{Space Telescope Science Institute, 3700 San Martin Drive, Baltimore, MD 21218, USA}

\author[0000-0002-1872-5398]{James Munday}
\affiliation{Institute for Physics and Astronomy, University of Potsdam, Karl-Liebknecht-Str. 24/25, 14476 Potsdam, Germany}
\affiliation{Department of Physics, University of Warwick, Gibbet Hill Road, Coventry CV4 7AL, United Kingdom}

\author[0000-0001-9873-0121]{Pier-Emmanuel Tremblay}
\affiliation{Department of Physics, University of Warwick, Gibbet Hill Road, Coventry CV4 7AL, United Kingdom}

\author[0000-0001-6566-7568]{Michael Abdul-Masih}
\affiliation{Instituto de Astrof\'isica de Canarias,
E-38205 La Laguna,
Tenerife, 
Spain}
\affiliation{Departamento de Astrof\'isica,
Universidad de La Laguna,
E-38206 La Laguna,
Tenerife, 
Spain}

\author[0000-0003-4200-7852]{Matthias Fabry}
\affiliation{Department of Astrophysics and Planetary Science,
Villanova University,
800 East Lancaster Avenue,
Villanova, 
PA 19085, 
USA}

\author[0000-0003-4560-7925]{Joseph Giammarco}
\affiliation{Eastern University, Dept.~of Astronomy and Physics, 1300 Eagle Rd, St.~Davids, PA 19087}

\author[0000-0001-5473-856X]{Kelly M.~Hambleton}
\affiliation{Department of Astrophysics and Planetary Science,
Villanova University,
800 East Lancaster Avenue,
Villanova, 
PA 19085, 
USA}

\author[0000-0002-1355-5860]{Herbert Pablo}
\affiliation{American Association of Variable Star Observers, 49 Bay State Road, Cambridge, MA 02138, USA}

\author[0000-0002-3051-274X]{Marcin Wrona}
\affiliation{Department of Astrophysics and Planetary Science,
Villanova University,
800 East Lancaster Avenue,
Villanova, 
PA 19085, 
USA}

\author[0000-0002-1913-0281]{Andrej Pr\v{s}a}
\affiliation{Department of Astrophysics and Planetary Science,
Villanova University,
800 East Lancaster Avenue,
Villanova, 
PA 19085, 
USA}

\begin{abstract}
Models of eclipsing binaries require the assignment of appropriate emergent intensities to the surface elements of the binary components. For distance-dependent modelling of flux-calibrated light curves, this necessitates an approximation of the absolute normal intensities of both components of the binary, as well as how their brightness varies across the stellar disks (limb darkening). Such surface intensities are often inferred from other physical properties of the synthetic binary (effective temperature, surface gravity, etc.) through the use of model atmospheres, which in turn are generally suited to a particular range of stellar types or parameters. Here, we present the major developments included in the PHOEBE 2.5 release (publicly available from \url{http://phoebe-project.org}), which improve the fidelity of model binaries comprising hot, compact stars.  These developments include the incorporation model atmospheres produced using the T\"ubingen Model Atmosphere Package (TMAP) and Montr\'eal/Tremblay codes (complementing the already incorporated PHOENIX and Castelli \& Kurucz models, primarily suited to main sequences stars and low-temperature giants). Similarly, PHOEBE v2.5 now allows for blending/extrapolation of model atmospheres, meaning one can continue to make use of model atmospheres in cases when a small number of surface elements have parameters outside the model atmosphere grid.  As an added value product, we also present tables of limb-darkening coefficients derived from the newly incorporated model atmospheres, such that they can be used as inputs in other binary modelling codes.
\end{abstract}



\section{Introduction} \label{sec:intro}

An essential element in all binary light curve modelling codes is the assignment of an appropriate emergent intensity to all visible surface elements of the discretized mesh which describes the geometry of the two stars, such that (when integrated) the total light from the binary can be modelled \citep{wilson71,prsa18}.  In some codes, the emergent intensities are derived via a user-provided surface brightness ratio in the passband of interest (e.g., JKTEBOP, \citealt{jktebop}, or ELLC, \citealt{ellc}), while others (including PHOEBE, \citealt{phoebe2}) make use of model atmospheres.  Here, the passband-integrated intensities are derived based on synthetic spectra created using model atmospheres, and then tabulated as a function of the defining model parameters   \citep[typically effective temperature, surface gravity and chemical composition;][]{vanhamme03}.  The resulting tables can then be interpolated to derive emergent intensities for a given surface element based on its model properties -- again usually the temperature (accounting for irradiation, gravity brightening, etc.), surface gravity and metallicity.  This approach has the added benefit of allowing the model light curves to be calculated in absolute units \citep[taking into account the distance and interstellar extinction of the system;][]{phoebe4}, which can then be directly compared to flux-calibrated light curves. However, the model atmospheres used (as well as the codes used to synthesise spectra from these models) are all subject to their own underlying assumptions and, as such, are generally valid only for a limited parameter space.  Outside of these ranges, many codes default to blackbody atmospheres which are, at best, a poor approximation \citep{phoebe2}.

PHOEBE was initially developed to make use of synthetic spectra produced with \citet[][referred to in PHOEBE as CK2004]{ck2004} model atmospheres which assume Local Thermodynamic Equilibrium (LTE) -- an assumption which breaks down for both the hottest and coolest of stars.  In \citet{phoebe4}, non-LTE PHOENIX atmospheres were incorporated (v2.2 release), extending the useful range of atmospheres to both lower temperatures and lower metallicities.  In this paper, we present the implementation of plane-parallel model atmospheres from the T\"ubingen non-LTE model atmosphere package \citep[TMAP;][]{rauch03,werner03} and Montr\'eal/Tremblay group \citep{tremblay09,tremblay13}.  These model atmospheres dramatically increase the fidelity of PHOEBE-computed observables for hot and compact stars -- including hot white dwarf (WD) and subdwarf O stars.  Additionally, we outline several other functionalities implemented in the PHOEBE code as part of the v2.5 release.  These include: Blending \citep[sometimes referred to as ramping;][]{vanhamme03}, Doppler boosting/beaming and user-defined features.  Blending allows for smooth transition from model atmospheres to black bodies, in order to avoid the need to fall back on the (demonstrably poor) blackbody approximation for entire stars when only a handful of surface elements lie outside the parameter range covered by model atmospheres (for example, due to the effects of irradiation or tidal distortion).  Doppler boosting was previously disabled in PHOEBE due to concerns over the way boosting factors were derived and tabulated - however the underlying functionality was never in question. Thus Doppler boosting is now reactivated allowing for users to provide factors calculated outside of PHOEBE.  User-defined features allow the user to modify models after computation or modify component mesh quantities. Example applications of this functionality include differential rotation, migrating spots and variable third light. 

\section{Hot, compact star atmospheres} \label{sec:tmap}

\subsection{TMAP Implementation}

Angle-dependent specific intensities were calculated using TMAP assuming pure Hydrogen or Hydrogen-Helium atmospheres and following the four grids (sdO, DA, DAO, DO) presented in \citet{reindl16,reindl23}.  These grids are implemented as separate atmosphere options inside PHOEBE to avoid any possible issues with interpolation between the different models (although we will later demonstrate that their behaviour is consistent).  The coverage of the different model grids, in terms of effective temperature, surface gravity and surface abundances, are detailed in Table \ref{tab:tmap_atm}. 

\begin{deluxetable*}{ccccl}
\tablecaption{New atmosphere models implemented into PHOEBE.\label{tab:tmap_atm}}
\tablehead{\colhead{PHOEBE name} & \colhead{$\teff$} &  \colhead{$\logg$} &\colhead{$\mathrm{log}(\mathrm{He}/\mathrm{H})$} & Reference \\ 
& \colhead{kK} & \colhead{dex} &\colhead{dex} &} 
\startdata
tmap\_sdO & 40 -- 140 & 4.75 -- 6.5  &  $-$1.55 -- $-$0.42 & \citet{reindl16}\\
tmap\_DA & 20 -- 200 & 4.0 -- 9.5 & $-$10\tablenotemark{a} & \citet{reindl23}\\
tmap\_DO & 40 -- 200 & 6.0 -- 9.5 & 9.4 & \citet{reindl23}\\
tmap\_DAO & 40 -- 200 & 6.0 -- 9.5 & $-$5 -- 0 & \citet{reindl23}\\
\hline
tremblay & 3.75 -- 60 & 5.0 -- 9.0 & $-$10\tablenotemark{a} & \citet{tremblay09}\\
\enddata
\tablenotetext{a}{The implemented atmosphere models are actually pure Hydrogen but are labelled as $\mathrm{log}(\mathrm{He}/\mathrm{H})=-10$ in order to share the same definition of the PHOEBE `loghefrac' parameter between the different hot, compact star models}
\tablecomments{Models close to or above the Eddington limit are not included.}
\end{deluxetable*}

\begin{figure*}
\plotone{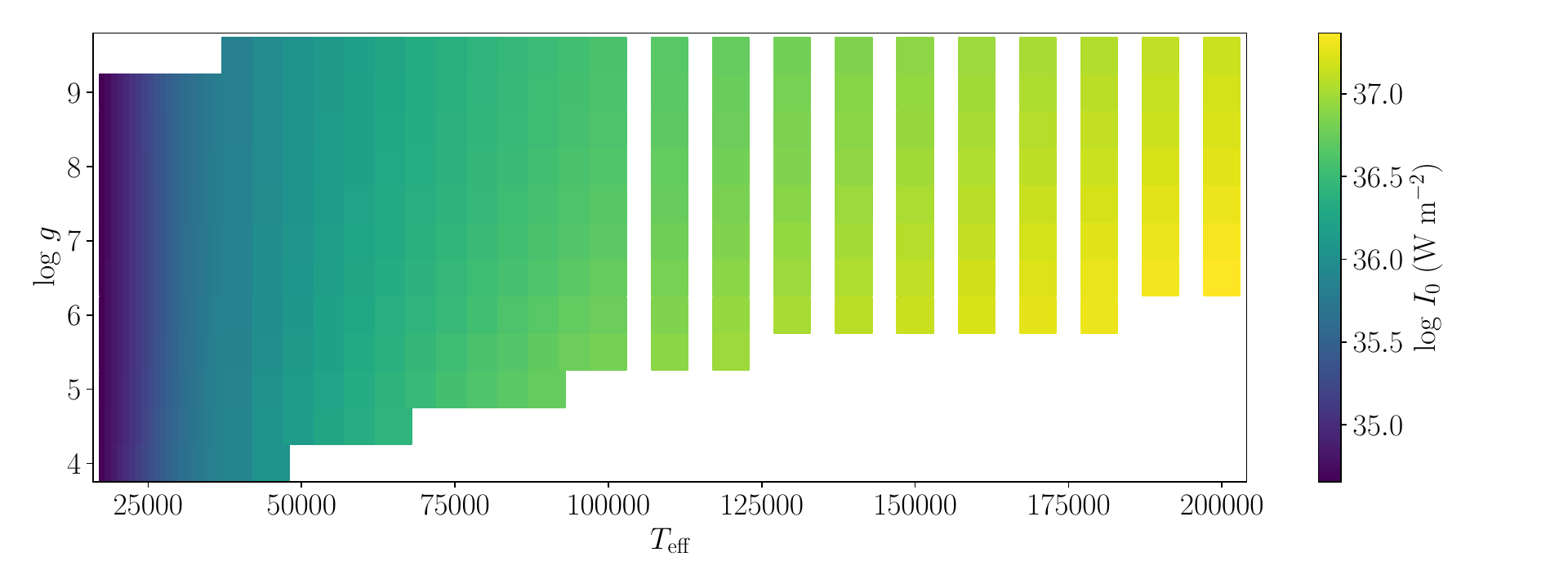}
\caption{The normal intensity in SDSS-$g^\prime$ band as a function of surface gravity and effective temperature for the TMAP DA WD models.  The white spaces are due to the sparser sampling of the grid for higher temperatures and regions where the models approach or exceed the Eddington limit. \label{fig:Inorm}}
\end{figure*}

The angles, across the stellar limb, for which emergent spectra can be calculated are hard-coded into TMAP, and cover 32 values of $\theta$ from 2.997\degr{} (close to the center of the stellar disk) to 89.922\degr{} (close to the edge of the stellar limb).  Just as for the other atmospheres incorporated into the code \citep[see][and references therein]{phoebe2,phoebe4}, PHOEBE's passband files contain tables of these TMAP angle-dependent spectral energy distributions multiplied by the passband transmission function to provide integrated emergent intensities as a function of the surface parameters -- effective temperature, surface gravity and abundance.  An example of the variation of normal intensity as a function of surface gravity and effective temperature for the DA atmospheres is shown in Fig.\ \ref{fig:Inorm}.

The major difference in implementation between the TMAP tables and those of the previously implemented atmospheres is the definition of the abundance (`abun' in PHOEBE). For CK2004 and PHOENIX, the abundance is defined as the metallicity, however this definition is not appropriate for the hot and compact stars covered by the TMAP grids.  As such, to avoid confusion with the `abun' parameter, the abundances of the TMAP grids specified using a new parameters, `loghefrac', defined as the logarithmic Helium to Hydrogen number fraction ($\mathrm{log}(\mathrm{He}/\mathrm{H})$). 

\subsection{Limb-darkening properties of the TMAP models}

\begin{figure*}
\plotone{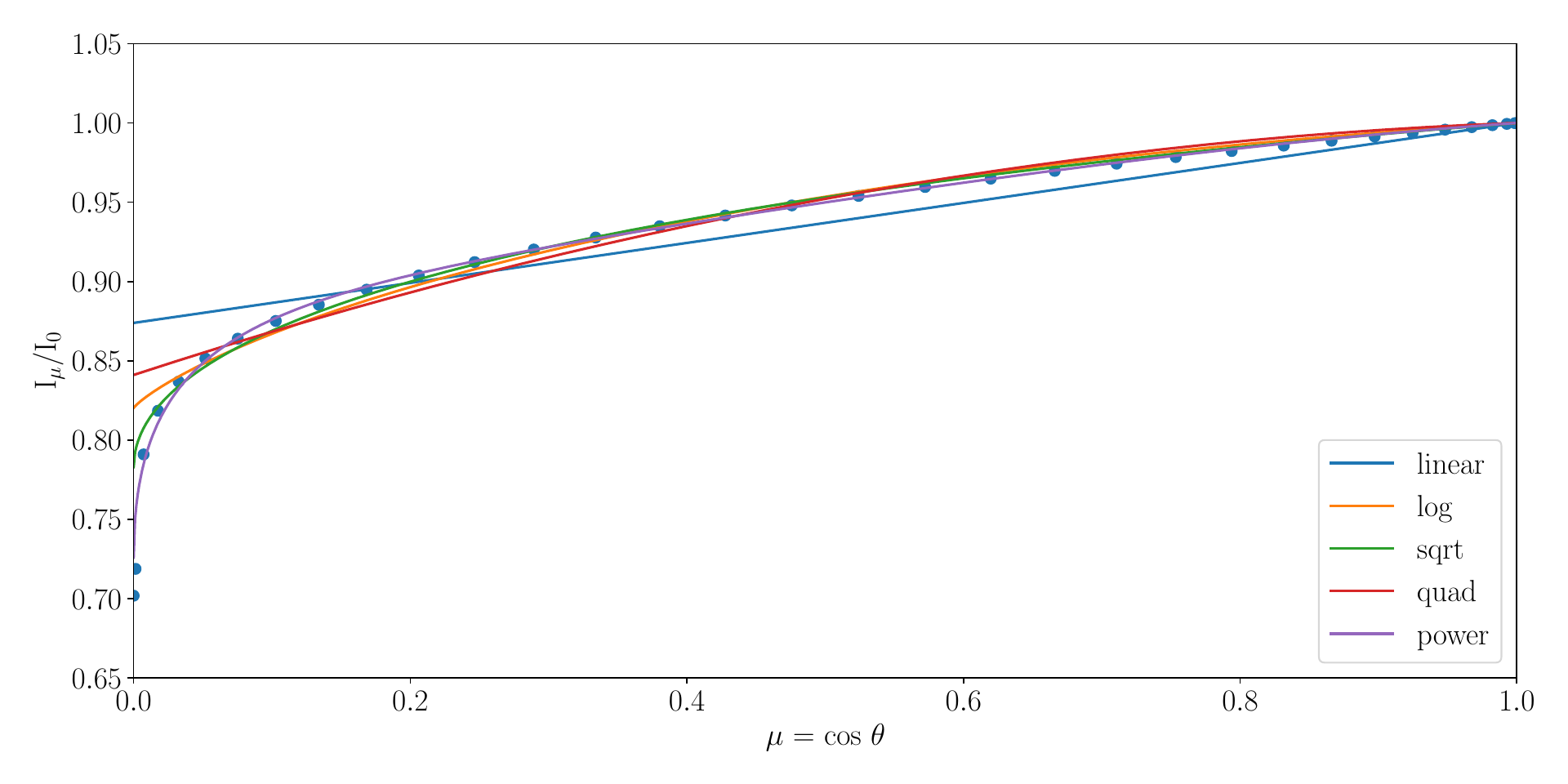}
\caption{The variation of emergent intensities in the SDSS-$g$ band as a function of angle across the stellar disk for a TMAP sdO model with $\teff=70$~kK, $\logg$=6.0 and a log (He/H)=$-0.97$.  The blue points show the TMAP sampling along the stellar disk, while the colored lines represent the fit to these points using different LD parameterisations. \label{fig:LDplot}}
\end{figure*}

An example of the variation in emergent intensity as a function of $\mu=\mathrm{cos}~\theta$ for a TMAP atmosphere in the $g$-band is presented in Fig. \ref{fig:LDplot}.  As an added value product, these angle-dependent passband intensities can also be used to derive limb-darkening (LD) coefficients for the standard parameterisations used in the literature \citep[see Sec.\ 5.2.3 of][]{phoebe2}\footnote{In \citet{phoebe2}, there is a typographic error in the definition of the logarithmic model where $log_{10}$ should, in fact, be the natural logarithm, $ln$ \citep[][]{klinglesmith70}.}.  While they are not required for PHOEBE \citep[which interpolates these tables on-the-fly at run time, thus avoiding the need to choose a parameterised approximation;][]{phoebe2}, they may be useful for other codes\footnote{PHOEBE can also be used to interpolate these tables of limb-darkening coefficients for use in other codes should the user wish.} and, as such, we provide machine-readable versions of the tables of these coefficients for the standard SDSS and Johnson filter sets for DA, DO and DAO white dwarfs as well as sdO stars in a .tar.gz package.  These values are derived by non-linear least squares fitting to the angle-dependent emergent intensities via the Levenberg-Marquardt algorithm \citep{lm}. 

\subsection{Consistency of the four TMAP grids}

\begin{figure*}
\plottwo{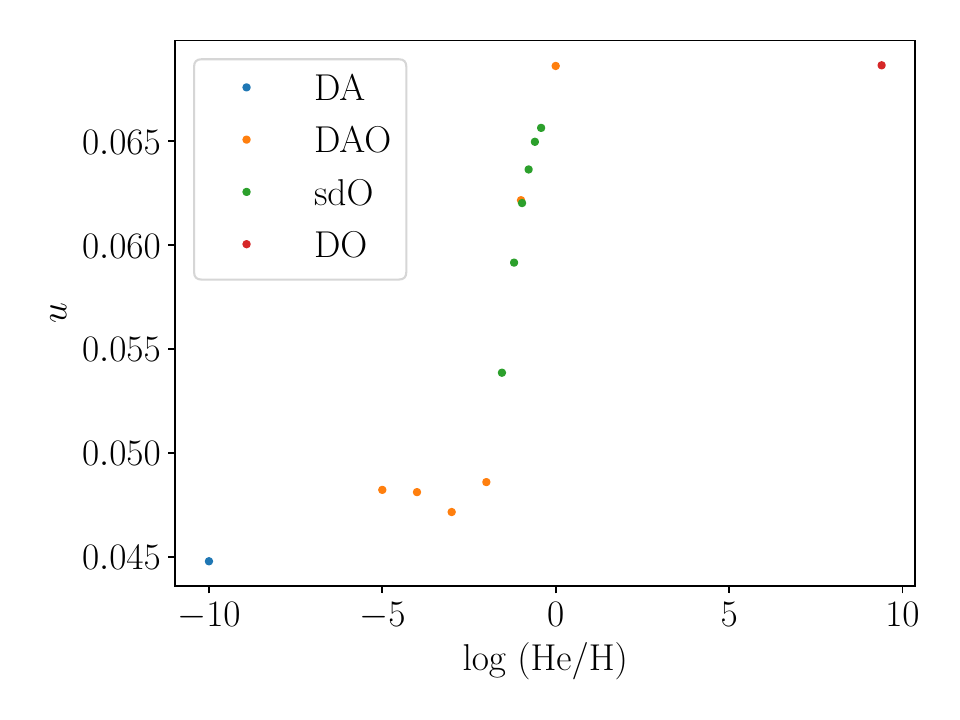}{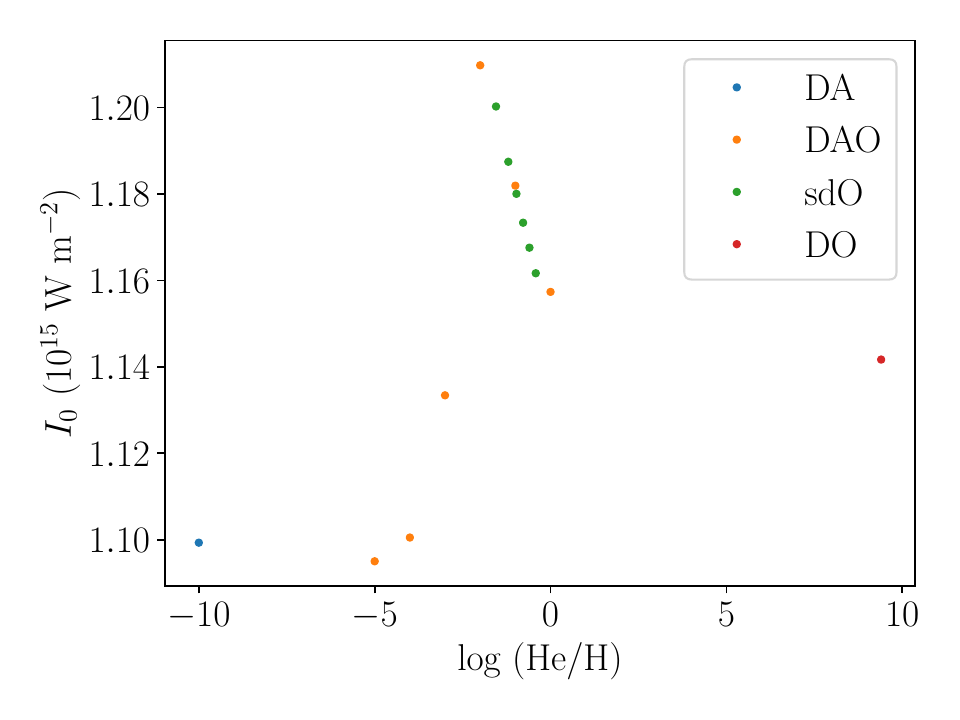}
\caption{The SDSS  g$^\prime$ band linear limb-darkening coefficient ($u$, left) and normal intensity ($I_0$, right) of the four TMAP model grids versus the log(He/H) abundance for models with effective temperature of 80kK and surface gravity of log $g=6.5$. \label{fig:TMAPabuns}}
\end{figure*}

The four TMAP model grids provide irregular coverage of the parameter space - particularly in terms of the abundance - however there are overlaps, specifically between the sdO and DAO grids.  Thus, we explore the key properties of the models (in terms of binary modelling) in order to test for consistency between the different grids.  To do this, we compare the normal intensities and derived linear limb-darkening coefficients as a function of abundance (He to H fraction by number) for a given combination of effective temperature and surface gravity in the range covered by all four grids (i.e. 40~kK $\leq \teff \leq 140$~kK and $6.0 \leq \logg \leq 6.5$).  An example, for $\teff=80$~kK and $\logg=6.5$,  is shown in Fig.~\ref{fig:TMAPabuns} demonstrating the continuous nature of the variation of limb-darkening coefficient and normal intensity as a function of abundance.

\subsection{Montr\'eal/Tremblay atmospheres}

The only other tables of limb-darkening parameters covering a similar range as those of our TMAP models were published by \citet[covering $\teff \leq$ 100~kK and 5.0 $\leq$ log $g$ $\leq$ 9.5]{claret20}.  They present two overlapping grids for DA white dwarfs comprising non-LTE TLUSTY models \citep{tlusty} extending to higher temperatures and LTE Montr\'eal/Tremblay models reaching down to lower temperatures \citep{tremblay09}.  The latter models are now also available as an atmosphere option in PHOEBE, with the implementation based upon the same grid of model spectra employed by \citet{claret20}.  For further details of the underlying models, we therefore refer the reader to \citet{claret20}.  We do not incorporate the TLUSTY models of \citet{claret20} into PHOEBE as an error in their calculation leads to erroneous intensities and limb-darkening coefficients\footnote{It is important to emphasise that this is not an error in TLUSTY or reflective of all synthetic spectra calculated using TLUSTY. Extensive testing and comparison between TMAP and TLUSTY has found that the two codes agree to within 1\% over an
extended wavelength range \citep{bohlin20}. In this case, it is simply an error in the intensities presented by \citet{claret20}.} (Tremblay, priv. comm.).

\subsection{Comparison of TMAP and Tremblay limb-darkening and normal intensities}

\begin{figure}
\plotone{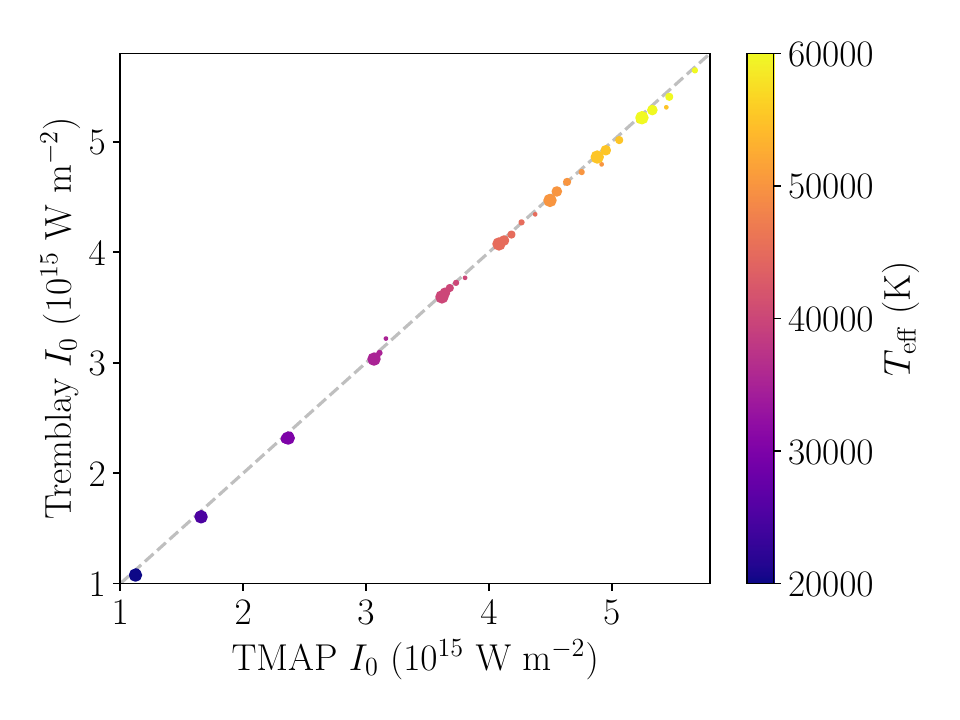}
\caption{The normal intensities of the TMAP and Tremblay DA models in the SDSS  g$^\prime$ band, where the point size is proportional to the surface gravity and the color represents the effective temperature.  The dashed line demarcates a perfect agreement between the models. \label{fig:Inorm_comp}}
\end{figure}

As a first step, we check the consistency between the TMAP DA and Tremblay models as implemented in PHOEBE, contrasting the normal intensities in the SDSS g$^\prime$ passband of both models as a function of effective temperature and surface gravity in Fig.\ \ref{fig:Inorm_comp}.  The agreement between the two models is clear, with the ratio of the normal intensities being within one standard deviation of unity (1.01$\pm$0.10 for SDSS g$^\prime$).

\citet{claret20} provides the normal intensities for their models as mW~m$^{-2}$~Hz$^{-1}$ requiring a passband (and thus spectrum) dependent conversion to W~m$^{-2}$ for direct comparison with the models implemented in PHOEBE\footnote{According to the notes added to the VizieR catalogues, the \citet{claret20} tables also require the application of an additional factor $\times$2 which are taken into account in the PHOEBE implementation of the Tremblay models}. For this, we assume that the effective wavelength of the passband is also the effective wavelength of the passband convolved with the emergent spectrum (i.e.\ the emergent spectrum is essentially flat in the range covered by the passband).  Under this assumption, the agreement between the Tremblay intensities tabulated by \citet{claret20} and the intensities in PHOEBE are good -- again being within one standard deviation of unity.

\begin{figure}
\plotone{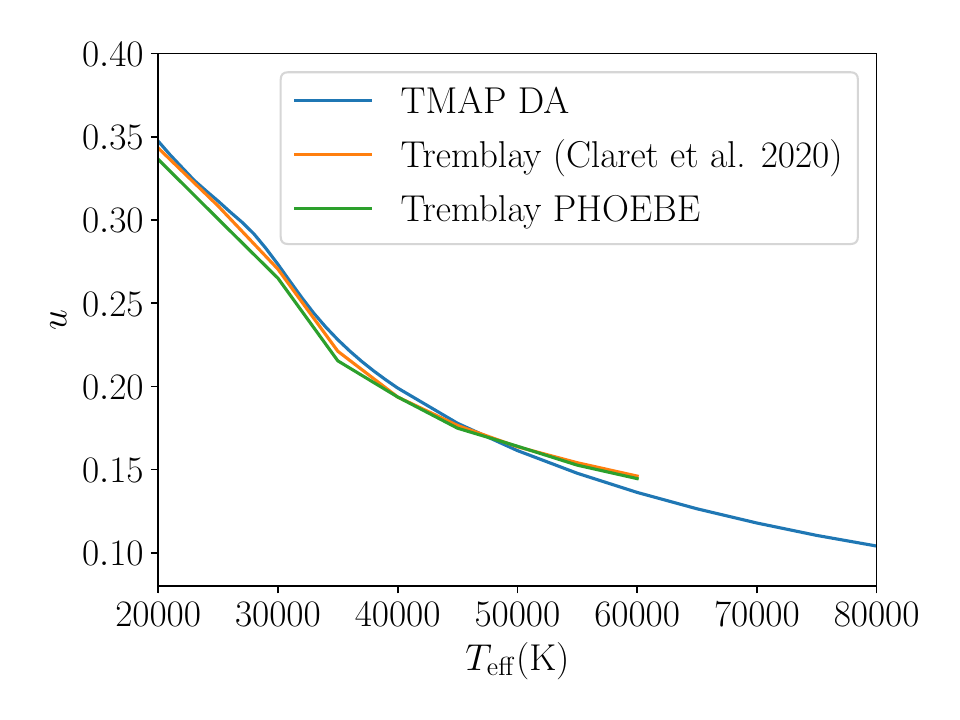}
\caption{The linear limb-darkening coefficient, $u$, as a function of effective temperature for DA models with log $g$=7.0. \label{fig:LinLD}}
\end{figure}

Comparing limb-darkening coefficients leads to a very similar conclusion.  There is generally very good agreement between the values measured by \citet{claret20} from the Tremblay models with the values from PHOEBE using the same models as well as using TMAP (see Fig.\ \ref{fig:LinLD}).  Small deviations are found between the Tremblay coefficients from \citet{claret20} and PHOEBE, principally due to the methodology used to fit the models.  \citet{claret20} interpolates the models to 100 equally spaced points along the stellar limb, while PHOEBE takes the 20 unevenly spaced points and weights them by the size of the interval between points.  The agreement between Tremblay and TMAP is similar to that found when comparing normal intensities.  

\section{TMAP/Tremblay versus blackbody atmospheres}

It has been clear for some time that approximating hot, compact stars as black bodies can lead to significant issues, for example, when considering the central ionising sources of photoionised nebulae \citep[e.g.,][]{pottasch08}.  Here, we attempt to characterise the impact that this can have on a modelled binary system.

\begin{figure}
\plotone{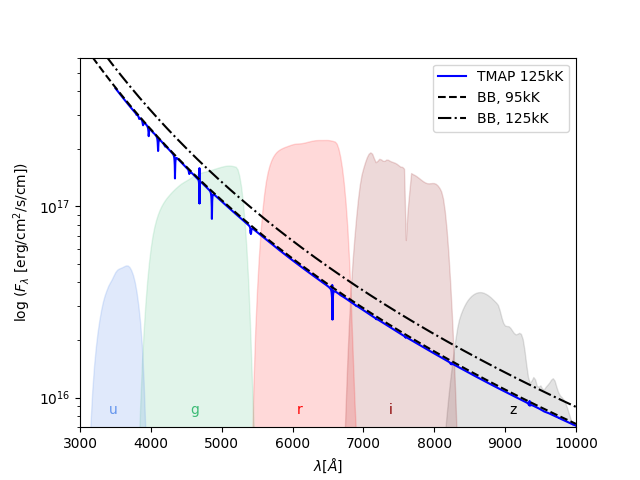}
\caption{A comparison of a 125~kK TMAP spectrum with blackbody spectra of different temperatures.  Standard SDSS filter bandpasses are underplotted to highlight that the differences are appreciable throughout the optical range. \label{fig:TMAPvsBB_spec}}
\end{figure}

Across the optical, TMAP passband-integrated intensities are approximately 25\% lower than for a blackbody of the same temperature \citep[see Fig.~\ref{fig:TMAPvsBB_spec} and, e.g.,][]{rauch03b}, depending on the passband, effective temperature, surface gravity and abundance (in some instances the difference is as large as 40\%!).  The principal impact of which is that for a given passband the temperature of a hot, compact star is likely to be  underestimated when approximated by a blackbody in a model light curve.

\begin{figure*}
\plotone{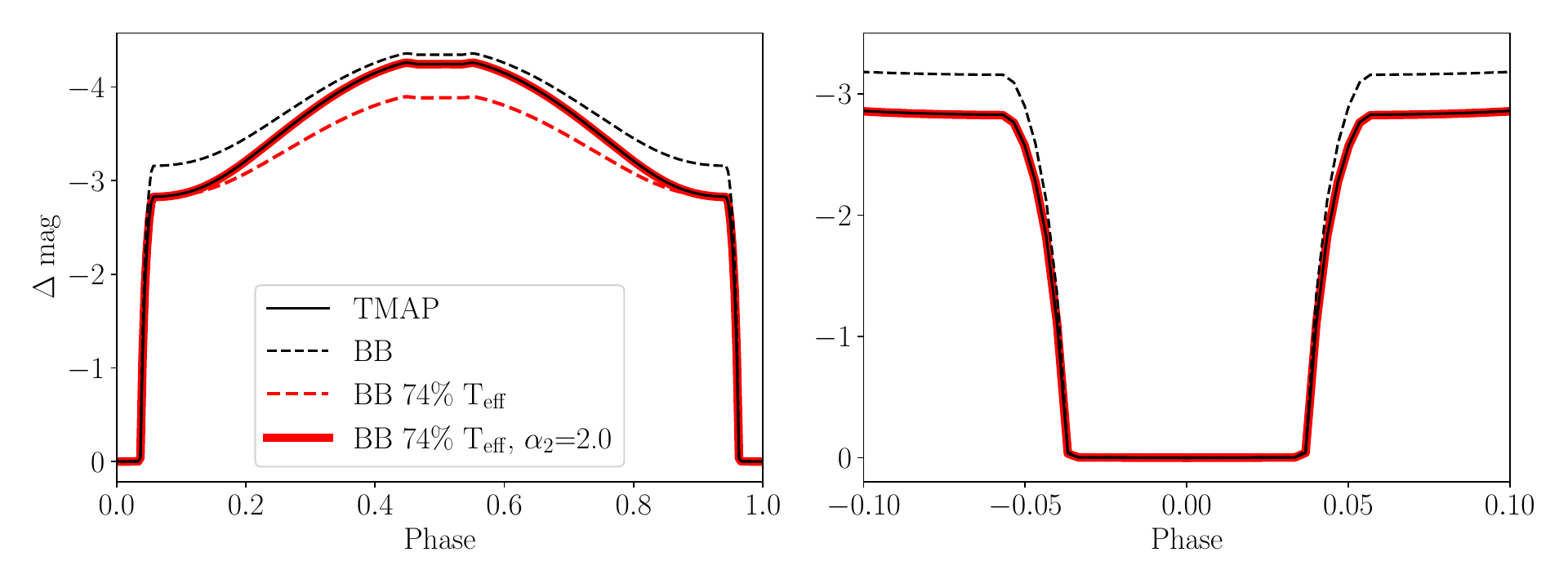}
\caption{A synthetic binary comprising a pre-WD ($\teff=127$~kK with TMAP atmospheres) in an 8-hr orbit with a K5V companion (solid black curve; TMAP).  For comparison, the same model was run with a blackbody atmosphere for the pre-WD (dashed black line; BB), with a blackbody atmosphere and a reduced temperature ($\teff\sim94$~kK, dashed red line; labelled ``BB  74\% $\teff$'') and with a blackbody atmosphere of reduced temperature and an increased secondary albedo (sold red line; labelled ``BB  74\% $\teff$, $\alpha_2=2.0$'').  \label{fig:TMAPvsBB_LC}}
\end{figure*}

When such hot and compact stars, like those covered by the TMAP model grids, are found in close binaries with lower temperature companions, they are often found to irradiate the face of the companion closest to them leading to a roughly sinusoidal variation as a function of orbital phase -- referred to as an irradiation or reflection effect \citep{horvat19}.  The shape and amplitude of this effect is a complex function of the passband and the binary parameters \citep[e.g.,][]{alencar99,jones17} but perhaps most critically, at least in the context of the comparison between TMAP and blackbody atmospheres, upon the luminosity of the irradiating star and the bolometric albedo of the irradiated star. As mentioned above, the difference in optical intensities between TMAP and blackbody atmospheres will often lead to underestimated temperatures for hot stars when modelling optical light curves -- this would also imply a reduced luminosity and thus reduce the level of irradiation experienced by a close companion.  Therefore, in order to reproduce the amplitude of an irradiation effect using blackbody atmospheres for the irradiating star, the bolometric albedo of the irradiated companion will need to be increased -- in many cases leading to highly unphysical albedos well in excess of unity.

To demonstrate the issues highlighted above, we create a synthetic binary comprising a young, 127~kK pre-WD \citep[with mass, temperature and radius taken from the post-AGB evolutionary tracks of][]{mmmb16} in an 8-hr orbit with a K5V companion.  Model light curves of this synthetic binary in the SDSS-$i$ band, calculated using TMAP and blackbody atmospheres are shown in Fig.~\ref{fig:TMAPvsBB_LC}, clearly demonstrating that the blackbody atmosphere results in a brighter pre-WD (as evidenced by the model light curve being $\sim$0.3 magnitude brighter upon egress of the primary eclipse).  In order to roughly reproduce the TMAP model light curve with blackbody atmospheres, the temperature of the pre-WD needs to be reduced to approximately 94~kK ($\sim$74\% the original $\teff$ used in the TMAP model) in the blackbody model.  Similarly, in order to compensate the reduced luminosity of the pre-WD with its reduced effective temperature in the new blackbody model, the albedo of the companion needs to be significantly increased (from 0.6 to an unphysical 2.0!) in order to reproduce the amplitude of the irradiation effect.

\section{Blending} \label{sec:blending}

New model atmospheres also highlight another problem that PHOEBE models long struggled with: the functionality when atmospheric parameters go out of model atmosphere bounds. There are several common cases where this can become an issue, for example:
\begin{itemize}
\item a semi-detached binary that has a few surface elements near the inner Lagrange point that become too cool or too rarefied for model atmospheres
\item a hot compact star that irradiates its companion, thereby exceeding model atmosphere boundaries
\item a rapidly rotating star with equatorial surface gravity falling off the grid.    
\end{itemize} 
In these and similar circumstances, PHOEBE previously raised an error that the chosen model atmosphere could not be used. The question is whether this is indeed warranted: losing the benefits of the sophisticated model atmosphere on account of a few surface elements that might fall off the model atmosphere grid.

To overcome this limitation, PHOEBE now supports two modes of operation that allows for a more graceful treatment of model atmospheres: \emph{extrapolation} and \emph{blending}. We discuss both, along with the typical circumstances that warrant their use, below. The underlying principle is the construction of an incompletely populated $n$-dimensional grid of physical quantities (intensities, limb darkening coefficients, boosting coefficients, etc.) that stem from model atmospheres.

Consider a model atmosphere with three basic input parameters: effective temperature ($\teff$), surface gravity ($\logg$) and heavy metal abundance ($\abun$). These parameters determine basic axes in the 3-dimensional space. In principle, given the values of the 3 parameters and assuming that they correspond to a physically allowed circumstance, we could use a model atmosphere code to synthesize a spectral energy distribution (SED), then multiply it with the passband response function, and ultimately integrate it into an emergent passband intensity. This is computationally prohibitive given that there are as many combinations of atmospheric parameters as there are surface elements on the star. Instead, each model atmosphere is sampled in discrete combinations of $\teff$, $\logg$ and $\abun$, the corresponding SEDs precomputed, and emergent passband intensities tabulated in a grid. As not all combinations of $\teff$, $\logg$, $\abun$ are physical, some grid nodes will remain undefined. Hence, we have constructed an incompletely populated $n$-dimensional grid of emergent intensities that PHOEBE can then use to linearly interpolate at any given 3-dimensional point $(\teff, \logg, \abun)$.

Linear interpolation on an incompletely populated grid is not a well posed problem. To demonstrate this, consider a 2-dimensional case with axes $a = (1, 2, 3)$ and $b=(4, 5, 6)$. This $3 \times 3$ grid requires 9 function values to be fully determined. Now suppose that a function value at $(2, 6)$ is unphysical or otherwise unavailable. How would interpolation be done in $(2.3, 5.3)$? We clearly have an over-dimensioned problem where any 4 nodes among the defined function values can be used. Given that function values (for example, intensities) are generally non-linear in basic parameters, the decision of what nodes to use for interpolation \emph{matters}. We elect to use the nearest \emph{fully defined hypercube}, in this case $(2, 3) \times (4, 5)$. The rationale, along with a more formal discussion of the interpolation strategy and the used terminology, is given in Appendix \ref{sec:ndpolator}.

The same construct can be used for linear extrapolation. This, too, is not a well posed problem. Extrapolating to an off-grid value again requires decisions on what defined nodes to use, which again \emph{matters} because of the non-linearity of the function values. We discuss these considerations in Appendix \ref{sec:ndpolator} as well.

Here we will use a catch-all term \emph{ndpolant} to stand for either interpolant or extrapolant in $n$-D space, and \emph{ndpolation} as the generalized term for linear interpolation and extrapolation in $n$-D space. Pertinent to the discussion at hand is that \emph{there exists} an ndpolant that can be computed from an incomplete $n$-dimensional grid. 

\begin{figure*}
\plottwo{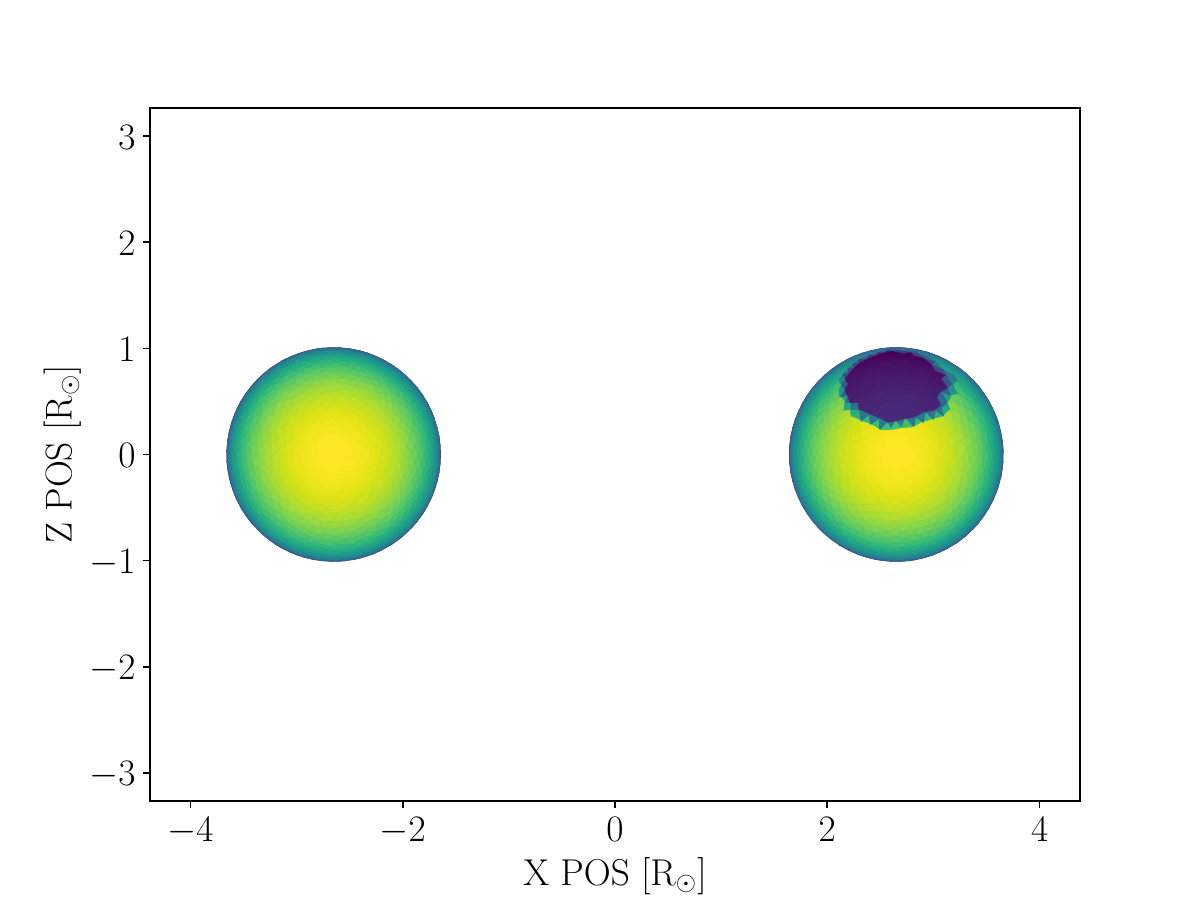}{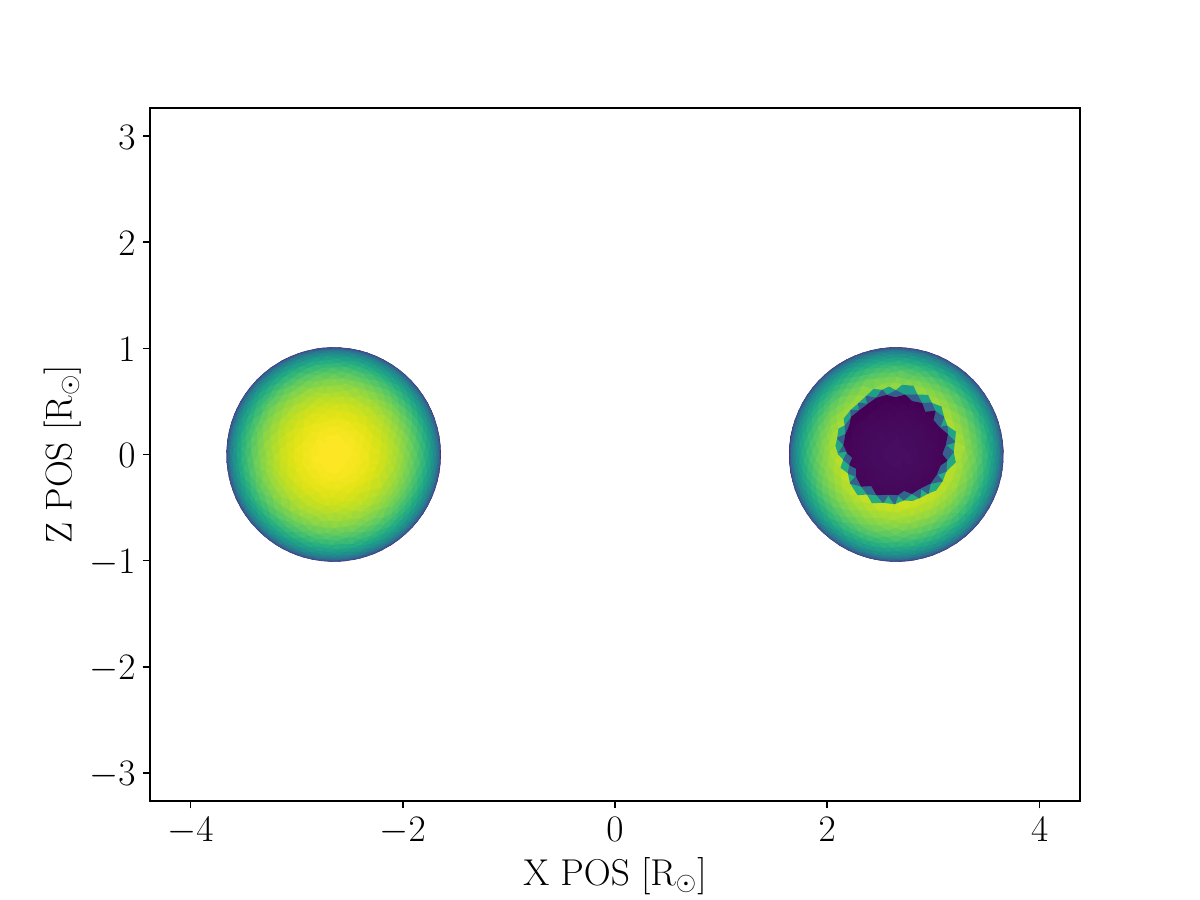}
\epsscale{0.5}
\plotone{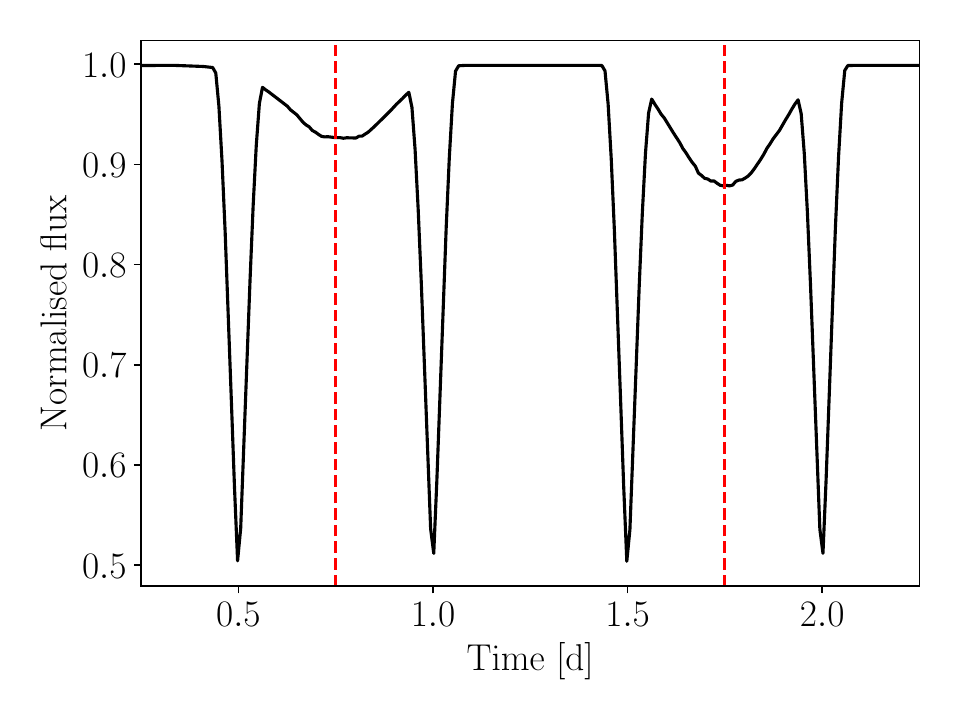}
\caption{A synthetic binary with a custom user-defined migrating spot.  The lower panel shows the synthetic light curve where the vertical dashed lines mark the times of the mesh plots shown in the upper panel. Both meshes are at the same orbital phase but the right-hand panel is one orbit later when the spot is closer to the line of sight, leading to a more pronounced change in flux as shown in the light curve. \label{fig:migrating_spot}}
\end{figure*}

Suppose that we want to evaluate a function value in an out-of-bounds query point $q = (\teff, \logg, \abun)$. PHOEBE supports two modes of operation:

\begin{description}
    \item[Extrapolation] driven by the PHOEBE parameter \texttt{extrapolation\_method}, which can be set to \texttt{`none'}, \texttt{`nearest'} or \texttt{`linear'}. If `none', extrapolation will not be used; this corresponds to the pre-2.5 functionality. The code will raise an error and stop execution. If `nearest', the nearest defined node will be used, where nearest is defined by the renormalized Euclidean distance (cf.~Appendix \ref{sec:ndpolator}). This is most appropriate for extrapolating limb darkening coefficients that are expressly non-linear. Finally, if 'linear', the function value will be linearly extrapolated from the nearest fully defined hypercube. This is most appropriate for extrapolating normal and specific emergent intensities.

    \item[Blending] driven by the PHOEBE parameter \texttt{blending\_method}, which can be set to \texttt{`none'} or \texttt{`blackbody'}. If 'none', the code will raise an error and stop execution. If `blackbody', the function value will be a blend between the extrapolated value from the given model atmosphere and the blackbody value, with the blending ratio dependent on the distance from the model atmosphere grid. This achieves a gradual transition from on-grid values, to extrapolated values in the immediate vicinity to the grid, that progressively morph into blackbody values as the distance from the grid increases. Blending is only appropriate for the normal emergent intensities as that is the only quantity inherent to the blackbody model atmosphere.
\end{description}

With this new functionality, PHOEBE can now compute observables even if some (or all) surface elements fall off the grid of supported model atmospheres, by either extrapolating or blending into blackbody atmospheres. Each such extrapolation or blending is stored in the mesh and presented to the user. This significantly facilitates the use of PHOEBE on the less well populated parts of the H-R diagram, and in systems with significant tidal and/or rotational distortion.

\section{Boosting} \label{sec:boosting}

Support for relativistic boosting (also frequently referred to as Doppler boosting or Doppler beaming) has been disabled in PHOEBE since the v2.2 release due to concerns over the way the boosting factors were derived from model atmospheres\footnote{For a complete discussion of the issue, please see \hyperlink{http://phoebe-project.org/docs/latest/tutorials/beaming_boosting}{http://phoebe-project.org/docs/latest/tutorials/beaming\_boosting}.}. However, as of v2.5 we return this functionality allowing the user to provide the boosting factors -- taking them from the literature \citep[e.g.,][]{claret20doppler} -- or fit for them. This approach has already been successfully used in, e.g., \citet{munday23}.

\section{User-defined features} \label{sec:features}

This release of PHOEBE also introduces the new capability to define custom code which can be executed at several different steps within PHOEBE's logic, including:

\begin{itemize}
    \item modifying the mesh before local value computations at any given model time;
    \item modifying local mesh values at any given model time: radial velocities, surface gravities, effective temperatures, and passband intensities; and
    \item modifying the completed synthetic model prior to evaluating the merit function.
\end{itemize}

In order to access this functionality, the user writes a class that inherits from a provided class in the PHOEBE, defines the input parameters, and optionally overrides any number of these ``hooks" into PHOEBE.  This class is then used to add a ``feature" to the PHOEBE system, in the same way as previously built-in features including spots.  The parameterization and code logic are then serialized along with the PHOEBE object (or saved file) and can be moved to another machine without having to move additional code or install extra dependencies.

This design allows for any custom code to be portable (including making use of external servers to offload expensive computations), shareable, and compatible with distributions and solver logic introduced in the previous release.  Any parameter written by the user is treated the same as a native parameter, enabling them to be optimized or sampled over simultaneously.

Example use-cases for these user-defined features might include: custom spot shapes and/or migration in time, differential rotation, simple intensity modulation from pulsations, or an additional time-dependent light contribution.

Figure \ref{fig:migrating_spot}, for example, shows a PHOEBE model with a spot which migrates in colatitude, having a more profound effect on the light curve as it approaches the orbital plane/line of sight.  In this case, the user writes a class which inherits from the default \texttt{Spot} class, defines the parametrization of the time-dependence (in this case the rate at which the spot's colatitude changes), and overloads the instantaneous position method which tells the existing spot code which triangles need their temperatures modified at any given time. The code for this example and others can be found in the online PHOEBE documentation for the v2.5 release.

\section{Summary}

As of the version 2.5 release, PHOEBE includes improved support for hot and compact stars via the inclusion of TMAP model atmospheres as well as blending between blackbody and TMAP/Tremblay/PHOENIX/CK2004 atmospheres.  This means that it is no longer necessary to use the blackbody approximation for stars whose surface elements have parameters which lie close to those covered by the incorporated model atmosphere grids.  Both of these improvements can have a dramatic impact on the stellar parameters derived by modelling light curves, as evidenced using the TMAP atmospheres where modelling with a blackbody can underestimate the true effective temperature by as much as $\sim$25\%.  Support for user-provided Doppler boosting coefficients has also been implemented (following the decision to deprecate interpolated boosting in PHOEBE's v2.2 release),  returning important functionality for cases where this effect can be appreciable (i.e.\ compact binaries).  New capabilities have also been implemented that allow user-defined features, which can modify the PHOEBE mesh, local values of the mesh or the completed synthetic model.

\begin{acknowledgments}
The development of PHOEBE is possible through the NSF AAG grants \#1517474 and \#1909109 and NASA 17-ADAP17-68, which we gratefully acknowledge.

DJ acknowledges support from the Agencia Estatal de Investigaci\'on del Ministerio de Ciencia, Innovaci\'on y Universidades (MCIU/AEI) under grant ``Nebulosas planetarias como clave para comprender la evoluci\'on de estrellas binarias'' and the European Regional Development Fund (ERDF) with reference PID-2022-136653NA-I00 (DOI:10.13039/501100011033). DJ also acknowledges support from the Agencia Estatal de Investigaci\'on del Ministerio de Ciencia, Innovaci\'on y Universidades (MCIU/AEI) under grant ``Revolucionando el conocimiento de la evoluci\'on de estrellas poco masivas'' and the European Union NextGenerationEU/PRTR with reference CNS2023-143910 (DOI:10.13039/501100011033). N.R. is supported by the Deutsche Forschungsgemeinschaft (DFG) through grant RE3915/2-1. P.-E.T. received funding from the European
Research Council under the European Union’s Horizon 2020 research and innovation programme number 101002408 (MOS100PC).
M.A. acknowledges support from the ``La Caixa'' Foundation (ID 100010434) under the fellowship code LCF/BQ/PI23/11970035.
\end{acknowledgments}

\software{Astropy \citep{astropy:2013, astropy:2018,astropy:2022}, 
        matplotlib \citep{2007CSE.....9...90H},
        numpy   \citep{numpy},
        PHOEBE \citep{phoebe2,phoebe3,phoebe4,phoebe5},
        scipy \citep{2020SciPy-NMeth},
        TMAP \citep{rauch03,werner03},
}


\newpage
\appendix

\section{$N$-dimensional interpolation on sparse grids} 
\label{sec:ndpolator}

Multi-variate ($n$-dimensional) interpolation and extrapolation are techniques used in mathematics, statistics and science to estimate unknown values between and beyond existing data points in a multi-dimensional space. Interpolation involves estimating the function value at points within the range determined by the existing data points, while extrapolation involves estimating the function value beyond that range. There are numerous robust implementations of multi-variate interpolation, including k nearest neighbors \citep{cover1967}, natural neighbor interpolation \citep{sibson1981}, radial basis functions \citep{hardy1971}, kriging \citep{cressie1990}, and many others. SciPy, for example, features an entire module for interpolation (\verb|scipy.interpolate|) that implements several multi-variate interpolation classes, including piecewise-linear, nearest neighbor, and radial basis function interpolators. Unfortunately, none of the implemented scipy methods lend themselves readily to extrapolation: at most they can fill the values off the convex hull with \verb|nan|s or a value supplied by the user. In addition, interpolators that operate on a regular $n$-dimensional grid do not allow any missing data; those values either need to be imputed by using unstructured data interpolators, or structured data interpolators need to be abandoned altogether.

Ndpolator (\url{https://github.com/aprsa/ndpolator}) aims to fill this gap: it can both interpolate and extrapolate function values within and beyond the grid definition range, and it can operate on incomplete grids. As a side benefit, ndpolator can estimate both scalar and vector function values, and it can reduce grid dimensionality for points of interest that lie on grid axes. It is optimized for speed and portability (the backend is written in C), and it also features a python wrapper. Given the gap in the multi-variate interpolation and extrapolation landscape, ndpolator development has been separated from PHOEBE and made available to the community as a standalone package.

\subsection{Ndpolator's operation principles}

Consider a scalar or a vector field $\mathbf{F}$ that is sampled in a set of $N$ $d$-dimensional points, $\mathbf F (x_1, \dots, x_d)_k$, $k = 1 \dots N$. Let these function values be sampled on a grid, where axes $\mathbf{a}_k$ span each grid dimension, so that $\mathsf{X}_{k=1}^{d} \mathbf a_k \equiv \mathbf{a}_1 \times \mathbf{a}_2 \times \dots \times \mathbf{a}_d$ is a cartesian product that spans the grid. Axis spacing need not be uniform: vertices can be separated by any amount that is required to make the grid sufficiently locally linear. If the grid is complete, i.e.~if there is a function value $\mathbf F(x_1, \dots, x_d)$ associated with each grid point, we have $N_c = \prod_k l(\mathbf a_k)$ function value samples, where $l(\mathbf a)$ is the length of axis $\mathbf a$. If the grid is incomplete, i.e. if some function values are missing, then $N < N_c$. Ndpolator defines grid points with sampled values as \emph{nodes}, and grid points with missing values as \emph{voids}. The points in which we want to estimate the function value are called \emph{query points} or \emph{points of interest}. The smallest $d$-dimensional subgrid that encloses (or is adjacent to) the query point is called a \emph{hypercube}.

\subsection{Unit hypercube transformation}

The first, most fundamental principle of ndpolator is that all interpolation and extrapolation is done on \emph{unit hypercubes}. In real-world applications, it is rarely true that all axes are defined on a unit interval. This can lead to vertices of significantly different orders of magnitude along individual axes. To that end, ndpolator first normalizes the hypercubes by transforming them to unit hypercubes: given the sets of two consecutive axis values that span a hypercube, $(\mathbf a_{1,p}, \mathbf a_{1,p+1}) \times (\mathbf a_{2,q}, \mathbf a_{2,q+1}) \times \dots \times (\mathbf a_{d,t}, \mathbf a_{d,t+1})$, the unit transformation maps it to the $[0, 1] \times [0, 1] \times \dots \times [0, 1] \equiv [0, 1]^d$ hypercube. All query points are subjected to the same transformation, creating unit-normalized query points. Therefore, interpolation (and extrapolation) \emph{always} operates on unit hypercubes, which is computationally the least expensive and numerically the most stable process. An additional benefit of this transformation is that extrapolation inherits the nearest hypercube's grid spacing, thus naturally accounting for (potentially) variable spacing in different regions of the grid.

\subsection{Sequential dimensionality reduction}

\begin{figure}[t]
    \centering
    \includegraphics[width=\linewidth]{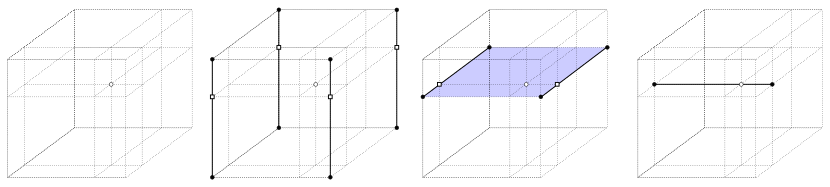} \\
    \caption{An example of sequential dimensionality reduction in 3 dimensions.}
    \label{fig:interpolation}
\end{figure}

The second operating principle of ndpolator is \emph{sequential dimensionality reduction}. Consider a 3-dimensional hypercube in Fig.~\ref{fig:interpolation}; let us assume that function values in all 8 corners of the hypercube are sampled, i.e. we have 8 nodes. The point of interest is depicted with an open symbol in the left panel, along with projections onto the hypercube faces. Ndpolator starts with the last axis, in this case $\mathbf a_3$, and it interpolates function values along that axis to the projections of the point of interest (second panel). These are \emph{univariate} interpolations. The process yields 4 vertices (depicted in open symbols), thereby reducing the initial dimension $d=3$ by 1, to $d-1=2$. The process is then repeated (third panel), this time along the second axis, $\mathbf a_2$, yielding 2 vertices, thereby reducing the dimension to 1. Finally, the last interpolation is done along axis $\mathbf a_1$ (right panel), yielding a single vertex, the point of interest itself. The dimension is thus reduced to 0, and the function value is determined. Thus, for an $d$-dimensional hypercube, ndpolator performs $\sum_{k=0}^{d-1} 2^k = 2^d - 1$ univariate interpolations to estimate $\mathbf F$ in the point of interest, which implies an $O(N)$ time complexity, $N = 2^d$ being the number of initial vertices within which one interpolates. This cost is acceptable for $d$ less than roughly 6 to 7. We note that if the interpolations along each of the dimensions are linear, the resulting value does not depend on the order in which the reductions were performed.

\subsection{Initial dimensionality reduction}

The third operating principle of ndpolator is \emph{initial dimensionality reduction}. In real-life applications it frequently happens that some of query point coordinates are aligned with the axes. For example, one of the axes might allow the variation of the second order variable, but its value usually defaults to the value that is sampled across the grid. When this happens, the initial hypercube dimension can be reduced by 1 for each aligned axis. The extreme case where the query point coincides with a node means that hypercube dimensionality is reduced to 0, and there is no need for interpolation. For that reason, ndpolator flags each coordinate of the query point as ``on-grid,'' ``on-vertex,'' or ``out-of-bounds.'' When ``on-vertex,'' hypercube dimension can be immediately reduced. When that happens, the time dependence is reduced by a factor of $2^m$, where $m$ is the number of coordinates aligned with the axes.

\subsection{Incomplete hypercubes}

The fourth operating principle of ndpolator is dealing with \emph{incomplete hypercubes}. If any of the hypercube corners are voids, we cannot interpolate. For that purpose, ndpolator keeps track of all fully defined $n$-dimensional hypercubes; when a query point lies within an incomplete hypercube, ndpolator finds the nearest fully defined hypercube and uses it to extrapolate the function value in the point of interest. While this is globally still considered interpolation as the query point is within the grid's definition range, the estimated function value is, strictly speaking, \emph{extrapolated} from the nearest fully defined hypercube. Note that, when grids are particularly sparse and functions strongly non-linear, that can cause a substantial accumulation of error. In such cases, unstructured interpolation techniques might be a better fit.

\subsection{Extrapolation modes}

The fifth operating principle of ndpolator is extrapolation. Ndpolator has three extrapolation methods: `none', `nearest' and `linear'. When extrapolation method is set to `none', the function value that is outside the range of axes is set to \verb|nan|. For extrapolation method `nearest`, ndpolator stores a list of all nodes and assigns a function value in the node that is nearest to the query point. Lastly, if extrapolation method is set to `linear', ndpolator linearly extrapolates from the nearest fully defined hypercube in a manner equivalent to dealing with incomplete hypercubes. The choice for extrapolation method depends on the multi-variate function that we are estimating; if it is highly non-linear, extrapolation should be avoided, so `none' and `nearest' might be appropriate; if it is largely linear or varies slowly, then a `linear' extrapolation method might be warranted. Ndpolator is a linear extrapolator, so it cannot adequately estimate non-linear multi-variate functions.

\subsection{Basic axes and associated axes}

The question of grid completeness is quite impactful for performance; that is why the sixth operating principle of ndpolator is to distinguish between \emph{basic} axes and \emph{associated} axes. Axes that can have voids in their cartesian products are referred to as \emph{basic}. For these axes, we need full ndpolator machinery to perform interpolation and extrapolation. On the other hand, a subset of axes may have all nodes in their cartesian products, i.e.~they are guaranteed to be sampled in all vertices that basic axes are sampled in; these are referred to as \emph{associated} axes. Given that their sampling is ascertained, interpolation and extrapolation can proceed without concerns for incomplete hypercubes -- that is, for as long as their basic hypercube counterparts (hypercubes spanned by basic axes) are complete. Each associated axis reduces the dimensionality of the hypercubes that need to be stored for extrapolation lookup, thus optimizing performance further.

\subsection{Function value dimensionality}

The seventh and final operating principle concerns \emph{function value dimensionality}. Most interpolators assume that the function value $\mathbf F$ is a scalar; ndpolator does not make that assumption. $\mathbf F_r(x_1, \dots, x_d)$ can be a scalar or a vector or arbitrary length $R$ (within reason, of course). It is then a requirement that all nodes are also $R$-dimensional. Ndpolator will then interpolate and extrapolate all function value components separately, and yield an $R$-dimensional estimate of the function value $\mathbf F$ in the point of interest.

\bibliography{phoebe6}{}
\bibliographystyle{aasjournal}



\end{document}